# AI AND NON AI ASSESSMENTS FOR DEMENTIA

Technical Report

Mahboobeh Parsapoor (Mah Parsa) and Hamed Ghodrati {mah.parsa, hamed.ghodrati}@crim.ca
Vincenzo Dentamaro {vincenzo.dentamaro}@uniba.it
Christopher R. Madan {christopher.madan}@nottingham.ac.uk
Ioulietta Lazarou, Spiros Nikolopoulos and Ioannis Kompatsiaris
{iouliettalaz, nikolopo, ikom}@iti.gr

## ABSTRACT

Current progress in the artificial intelligence domain has led to the development of various types of AI-powered dementia assessments, which can be employed to identify patients at the early stage of dementia. It can revolutionize the dementia care settings. It is essential that the medical community be aware of various AI assessments and choose them considering their degrees of validity, efficiency, practicality, reliability, and accuracy concerning the early identification of patients with dementia (PwD). On the other hand, AI developers should be informed about various non-AI assessments as well as recently developed AI assessments. Thus, this paper, which can be readable by both clinicians and AI engineers, fills the gap in the literature in explaining the existing solutions for the recognition of dementia to clinicians, as well as the techniques used and the most widespread dementia datasets to AI engineers. It follows a review of papers on AI and non-AI assessments for dementia to provide valuable information about various dementia assessments for both the AI and medical communities. The discussion and conclusion highlight the most prominent research directions and the maturity of existing solutions.

# 1 Introduction

Dementia refers to a gradual decline in cognitive ability that leads to various impairments [1, 2], including memory and language deficits [3] in people, mainly older adults [4, 5]. The World Health Organization (WHO)'s reports stated that dementia could be a global mental health problem (see Figure 1), and by 2040, around 81 million people across the world will be affected by dementia. Thus, identifying older adults with dementia is essential for helping them be prepared for the future [6]. Furthermore, they can start using medical and non-medical treatments to slow down the progress of cognitive decline. However, since there is no single clinical assessment to accurately and quickly identify older adults who might be at the risk of dementia or at the early stage of dementia [7, 8], clinicians might use a set of various clinical assessment methods (CAMs) (including cognitive assessments and blood tests) for dementia. So, they face challenges in selecting appropriate assessments with the capability to detect cognitive impairments, behavioral disturbances, and functional impairments associated with stages of dementia or sub-types of dementia. In the era of AI and considering the recent breakthroughs in AI applications for healthcare, particularly dementia care, AI-based assessments can be considered as clinical assistants in identifying patients with dementia (PwD) [9, 10].

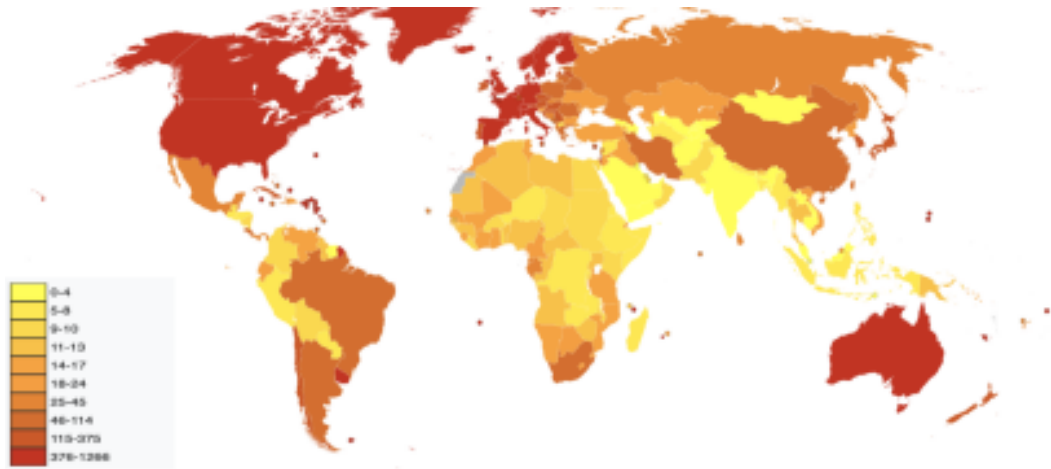

Figure 1: Death from dementia across the world.

Several articles focused on reviewing AI-powered systems for dementia. For example, the authors of [11] reviewed neuroimaging techniques of dementia. Another review about neuroimaging techniques for dementia has been done in [12]. As another example, the authors of [13] reviewed machine learning models applied to predict the progress of MCI to AD using neuroimaging data. As another instance, the authors of [14] provided a systematic review of using eye tracking technologies for detecting various types of cognitive disorders including dementia. The authors of [15] conducted a literature review on IoT wearable sensors and devices that have been used in elderly care settings. The authors of [16] conducted a systematic review of ML-based approaches for dementia prediction. In contrast to the above mentioned survey articles, this paper aims at providing helpful information about different types of AI and non-AI assessments for dementia. It seeks to answer the following questions: 1) What are the strengths and weaknesses of various clinical assessment methods for dementia?; 2) How can the AI community develop AI-based assessments for dementia?; 3) What are the challenges in developing AI-based assessments for dementia?; 4) What are the challenges in integrating AI-based assessments in dementia detection procedures?; 5) What are the challenges in integrating AI-based assessments in dementia detection procedures?; 6) What is the future of AI-based assessments for dementia?

# 2 Non-AI Assessments for Dementia

Clinicians and mental health professionals employ various types of clinical assessments to detect early signs of dementia or its subtypes in older adults. We use the term non-AI assessments to refer to them and provide an overview of diverse non-AI assessments mainly used to identify individuals with dementia in the primary care setting.

## 2.1 Dementia and its Different Subtypes

Dementia[1] is a progressive cognitive decline, which is caused by pathological changes in brain areas (e.g., the cerebral cortex). Different types of pathological changes can be associated with varying subtypes of dementia[2] (see Table 1). Alzheimer's disease (AD) is the most common type of dementia[3] and caused by cell damage [19], which can affect the brain regions including the hippocampus [20]. Vascular dementia (VaD)[4] is the second most common type of dementia. It can cause the impairment of memory and cognitive functioning resulting from cerebrovascular disease (CVD) [21]. Dementia with Lewy Bodies (DLB) is the third most common type of dementia. It is the result of the expansion of aggregated $\alpha$-synuclein protein in Lewy bodies and Lewy neurites (abnormal neurites in diseased neurons), which can cause severe damage to brain cells [22, 23] of the olfactory bulb, the dorsal motor nucleus of the vagal nerve, the peripheral autonomic nervous system, which encompasses the enteric nervous system, and the brainstem [24]. Frontotemporal dementia (FTD) or frontotemporal degenerations [25] has been defined as a combination of diseases caused by increasing nerve cell loss in the brain's frontal lobes and its temporal lobes [26]. Parkinson's disease dementia (PDD) is a type of dementia that can be experienced by patients living with Parkinson's [27]. PDD causes thinking and reasoning decline in patients. The main reason for PDD is the brain changes caused by Parkinson's disease, which mainly damages the motor cortex area of the brain.

Table 1: Subtypes of Dementia

| Dementia Subtypes | Clinical Symptoms | Pathological changes in brain area | Ref. |
| --- | --- | --- | --- |
| AD | Depression; Memory loss | Hippocampus | [28] |
| VaD | Executive dysfunction | Subcortical frontal | [21] |
| DLB | Gait and sleep problems | Cortical and subcortical; Olfactory bulb; Dorsal motor nucleus of the vagal nerve | [24, 29] |
| FTD | Behavioral, personality, movement and speech and language disorders | Frontal and temporal lobes | [26, 30, 31] |
| MD | Various symptoms | Various regions | [32] |
| PDD | Thinking and reasoning decline | Motor cortex | [27] |

---

[1] The Latin root of the word "Dementia" is "demens", "which means being out of one's mind" [17] or "a departure from previous mental functioning" [5].
[2] For example, problems including agnosia, spatial disorientation, language problems, apraxia, amnesia because of damage to the cerebral cortex, which is referred to as cortical dementia [18].
[3] Interested readers can refer to Glossary of Alzheimer's Disease Terms for getting more information about terminologies related to AD.
[4] Provokes troubles in the supply of blood to the brain [21].

## 2.2 Behavioral, Cognitive, Functional, Psychological and Neuropsychological Assessments of Dementia

There is no single tool to assess dementia [33, 34], thus, clinicians employ various types of tests and assessments including blood, urine tests, behavioral, cognitive [35, 36, 37], functional, psychological and neuropsychological assessments [35, 36, 37] to identify individuals with dementia [38]. In general, an ideal assessment should be valid (i.e., should show both face validity and construct validity), reliable (i.e., should show both inter-rater reliability and test–retest reliability), practical and objective (i.e., should show what exactly is normal and abnormal) [1].

One categorization of such assessments can be done on the basis of the types of abnormalities that they can detect. For example, behavioral assessment methods (BAMs) measure variations in behaviors and psychological tendencies in individuals. These changes are referred to as behavioral and psychological symptoms in dementia (BPSD) [1] and encompasses agitation, anxiety, elation, irritability, depression, delusions, hallucinations, psychosis, aggression, disinhibition, sleep deprivation and appetite disturbances [39, 40]. Another example is functional assessment methods (FAMs). FAMs measure the functional capacity of individuals by asking questionnaires from patients, patient's relatives, or caregivers [41] about the functional status of patients with dementia. Clinicians employ FAMs to diagnose the progressive loss in the ability to perform activities of daily living or functional disability [42]. Another category of assessments is neuropsychological assessment methods (NAMs), which are useful for: 1) Objectively assessing cognitive dysfunction; 2) Understanding which cognitive function is impaired; And 3) how seriously cognitive function has been impaired [43] at the early stage of dementia. Using NAMs, clinicians can successfully distinguish the severity of a patient's cognitive and behavioral impairments.

The main category of non-AI assessments is cognitive assessments methods (CoAMs), which can assess: 1) learning and memory, 2) language, 3) visuospatial, 4) executive and 5) psychomotor [44]. There are two main categories for CoAMs: comprehensive and non-comprehensive methods. The former examine all aspects of cognitive functioning, including orientation (e.g., patients are asked about the time and place), memory (e.g., patients are asked about recalling address), attention (patients are asked about months of the year and asked to spell a word forwards and backwards), executive functioning (e.g., patients are asked to do letter and category fluency tasks), language (e.g., patients are asked to name object), visuospatial and perceptual processing [18]. The latter evaluates only one cognitive status (e.g., memory or language).

Various CoAMs, which are mostly usable at primary care settings [45] have been listed in Table 2. Some examples of CoAMs are provided in the following: 1) Alzheimer's Disease Assessment Scale–Cognitive Subscale (ADAS-Cog) [46, 47] includes 11 items to evaluate attention, orientation, memory, language, visual perception, and visuospatial skills. In [48], an online version of the ADAS-Cog test has been referred to as computerized ADAS-Cog (cADAS-Cog) includes additional following tasks: delayed recall, number cancellation, and maze tasks.

2) Mini-Mental State Examination (MMSE) is an 11-item questionnaire that tests five cognitive functions including: orientation, attention, memory, language and visual-spatial [1, 33, 49, 50, 51, 52, 53, 54]. A standardized version of MMSE that is referred to as (SMMSE) has been developed to measure older people's cognitive function. Furthermore, the results of the SMMSE test along with the results of history and physical assessments can assist clinicians to differentiate between different subtypes of dementia [55]. Another version of MMSE is telephone-mini mental state examination (tMMSE), [56]). The MMSE tests have shown acceptable sensitivity to detect dementia with good inter-rater reliability. Running the MMSE test is quick and easy. The original MMSE test has been translated into different languages (i.e., there are more than 15 language versions of the MMSE [57]) [58]. The main disadvantage of MMSE is that clinicians cannot rely on that in detecting MCI and track the progress of dementia [58].

3) Montreal Cognitive Assessment (MoCA) consists of a 30-points scale and takes around 10 to 12 minutes [59, 60, 61] to be completed. It is a useful CoAM for different types of dementia such as AD, FTD and DLB, PDD but also for mental illnesses such as sleep behavior disorder. The online version of MoCA has been proposed as electronic Montreal Cognitive Assessment (eMoCA) [62], which could be useful to assess cognitive impairments in older adults that have not access to local clinical services. MOCA is a valid multi-language test [63] and several non-English versions of MoCA including French, Japanese, Chinese, Dutch, Portuguese, Turkish, Croatian, Sinhala, Arabic, Malay, Italian, Korean, Thai and so on have been developed. The MOCA can be used to distinguish patients with mild cognitive impairment (MCI) and mild dementia (MD) [63]. Other versions of MOCA are MoCA-V that has been developed for patients with vision impairment and MoCA-H that is useful for people with acquired hearing impairment [64].

Table 2: Non AI Assessments

| Name | Type | Characteristics | Ref. |
|---|---|---|---|
| Functional Independence Measure (FIM) | FAMs | Disability test; Not a test for diagnose | [65] |
| Frontal Assessment Battery (FAB) | FAMs | For detecting FTD; A memory and language test | [66] |
| Instrumental Activities of Daily Living (IADL) | FAMs | Multi-disease (can differentiate AD from MCI); Better than Barthel Index; High sensitivity | [67, 68] |
| ABC Dementia Scale (ABC-DS) | BAMs | Multi-stage detection; C-ABC[11] | [69, 70, 71, 72] |
| Behavioral Dyscontrol Scale (BDS) | BAMs | High accuracy; Multi-disease (can differentiate AD from MCI) | [73] |
| Dementia Mood Assessment Scale (DMAS) | BAMs | Detect depression and dementia; Reasonable validity; Reasonable inter rater reliability and intraclass correlation | [74, 75] |
| Hopkins Verbal Learning Test (HVLT) | BAMs, CoAMs | Multicultural; Quick (10 minutes); Effective test for both clinical and research settings | [76, 77, 78, 79] |
| Neuropsychiatric Inventory (NPI) | BAMs | Quick; Severity of delusions, agitation, depression, and behavioral irritability; Multi-disease detection (can differentiate AD from MCI) | [80, 81] |
| Relevant Outcome Scale for Alzheimer's Disease (ROSA) | BAMs | Multi-stage detection of AD | [82] |
| California Verbal Learning Test (CVLT) | NAMs | Memory and language test; CVLT-II and CVLT-III | [83] |
| Sentence Repetition Tasks (SRTs) | NAMs | Acceptable sensitivity; Not Sensitive to the educational level; ASRT | [84] |
| Teadinail Making Test (TMT) | NAMs | Most commonly-used | [36] |
| Token Test (TT) | NAMs | Comprehensive; Quick; Low cost | [85] |
| Semantic Verbal Fluency (SVF) test | NAMs | Language; A quick and easy-to-apply test; High sensitivity and specificity for the diagnosis of dementia; Detecting cognitive decline | [86] |
| Arizona Battery for Communication Disorders of Dementia (ABCD) | CoAMs | Comprehensive; Multi-age; Multi disease detection; Culturally sensitive | [87, 88] |
| Addenbrooke's Cognitive Assessment (ACE) | CoAMs | Multi-disease (differentiate AD from FTD); Higher accuracy than MMSE; Higher accuracy than MMSE; ACE-III has High sensitivity to early cognitive dysfunction, other version (ACE-R and ACE-III) | [89, 90] |
| Alzheimer's Disease Assessment Scale–Cognitive Subscale (ADAS-Cog) | CoAMs | Comprehensive; Multi-disease (an differentiate AD from FTD); cADAS Cog[5] | [46, 47, 48] |
| Eight-item Informant Interview to Differentiate Aging and Dementia (AD8) | CoAMs | Comprehensive; Quick ( 3 minutes); Easy; Culturally sensitive | [91] |
| Abbreviated mental test score (AMTS) | CoAMs | Quick; Easy; High specificity; Low sensitivity; not a "rule-out" test | [92, 93, 94] |
| Cambridge Assessment of Memory and Cognition (CAMCOG) | CoAMs | Comprehensive; Better that MMSE; Excellent sensitivity and specificity | [92, 95] |
| Cambridge Mental Disorders of the Elderly Examination (CAMDEX) | CoAMs | Comprehensive; High inter-rater reliable; High sensitivity | [96] |
| Clinical Dementia Rating (CDR) | CoAMs | Comprehensive; Online CDR; Modified CDR Test (Alzheimer's / Dementia Online Test – Questions ) | [97, 98, 99] |
| Clock-drawing Test (CDT) | CoAMs | Easy to run | [92, 100] |
| Codex (Cognitive Disorders Examination) | CoAMs | ML based assessment method (Based on decision tree); High sensitivity and specificity; Detection of dementia and MCI | [101] |
| General Practitioner Assessment of Cognition (GPCOG) | CoAMs | Reliable; Valid; Efficient; Quick ( 4 to 5 minutes); Easy; High sensitivity; High specificity | [102, 103] |
| Informant Questionnaire on Cognitive Decline in the Elderly (IQCODE) | CoAMs | Quick | [104, 105, 106, 107, 108] |
| Mini-Mental State Examination (MMSE) | CoAMs | Comprehensive; Easy; Acceptable sensitivity and specificity; Not a stand alone single administration method; Not multi level educational; sMMSE/ tMMSE | [33, 49, 1, 50, 51, 52, 53, 54, 56, 55] |

---

[5]It includes following tasks: Delayed Recall, Number Cancellation, and Maze tasks

| | | | |
|---|---|---|---|
| Months Backwards Test (MBT) or Months of the Year Backwards Test (MYBT) | CoAMs | Comprehensive test; Quick; Easy; Multi-stage of Dementia | [109, 110, 111] |
| Montreal Cognitive Assessment (MoCA) | CoAMs | Comprehensive test; Better than MMSE to detect early stage of dementia and MCI; Multi disease detection; eMoCA | [59, 60, 62] |
| Mini-Cog test | CoAMs | Comprehensive; Quick; Easy; Reason able sensitivity and specificity | [112] |
| Saint Louis University Mental Status (SLUMS) | CoAMs | Comprehensive; High sensitivity; High specificity; Better at detecting mild neurocognitive disorder, which cannot be detected by the MMSE; SLUMS-C | [113, 114, 115] |
| Six Item Cognitive Impairment Test (6-CIT) or the Six Item Orientation Memory Concentration Test | CoAMs | Good sensitivity and specificity; Easy to apply; Reliable; Valid; Comparable performance with MMSE | [92, 116, 51, 117] |

In addition to the above methods, some studies suggested using online assessment methods (see Table 3) to evaluate cognitive impairment in older adults. The main advantage of online assessment methods is that they can be used remotely to evaluate cognitive impairments in older adults who have not access to local clinical services [120, 121, 122].

### 2.2.1 Non AI Assessments for Distinguishing Different Types and Stages of Dementia

For individuals suspected of dementia, clinicians can start using assessments such as Mini-Cog or GPCOG [132] to distinguish the types of dementia. Based on the results of the above tests, they can do further evaluation using more in-depth assessments such as the MOCA, MMSE, and SLUMS along with laboratory tests and neuroimaging-based assessments. Among different assessments mentioned above, the MOCA test has been mentioned as a sensitive test to detect different subtypes of dementia at their earliest stages [133]. To identify patients with AD, clinicians can use an assessment from one of CoAMs, BAMs or NAMs [134] (BDS, IADL and SFT) or a combination of two assessments to evaluate memory and language impairment (i.e., the first symptoms of AD [135]) in individuals. These assessments can help them detect if an AD patient is at its earliest stage of the disease. Furthermore, ROSA can be used to detect patients with AD and find out the severity of the disease. To identify patients with FTD, clinicians can use ADAS-Cog and ACE. To diagnose a patient with VAD, clinicians need to consider neuroimaging-based assessments, and pathological tests based on three criteria: 1) Proof of dementia or cognitive impairment; 2) Proof of CVD; And 3) Proof of a causal relationship between cognitive impairment and CVD. For identifying a patient with DLB, Clinicians use MMSE and MOCA, CDR, CVLT-II. Also, tests that measure the processing speed and alternating attention such as "Stroop Tasks", "trail Making tasks", "phonemic fluency," and "computerized tasks of reaction time" can be used to diagnose patients with DLB. To detect different stages of dementia, MOCA, ABC-DS and MBT could be useful.

### 2.2.2 Challenges, Strengths and Weaknesses of using non-AI Assessments

Most assessments have failed to diagnose people with dementia with high sensitivity and specificity [1] and wrong diagnoses of the types of dementia or its severity could significantly affect the type of required treatments, the health insurance and social rights of individuals [136]. Thus, choosing appropriate assessments is challenging for clinicians. Another challenge is that most assessments mentioned above have been developed for assessing cognitive impairments or behavioral changes in native English speakers. Thus, they would need to be modified to be used to evaluate cognitive impairment in non-English speakers. Another problem is that most of the mentioned assessments are not usable to examine cognitive impairment in older adults with intellectual disabilities [137] or individuals with dual sense loss [138]. Furthermore, the majority of non-AI assessments need to be administered by specialists or trained clinicians; thus family members or caregivers cannot easily run them to identify older adults with dementia.

There is no doubt that each non-AI assessment has its strength and weakness, and many experimental works would be needed to discover those. However, in general, a robust assessments of dementia should: 1) Have high sensitivity and specificity; 2) Be quick; 3) Be easy to be administered (also a method that can be considered as a self-administered method such as SAGE); 4) Be usable in primary care settings; 5) Be useful for detecting different stages of dementia and different types of dementia; 6) Be applicable for individuals who are within different ages, coming from various cultures and having diverse educational backgrounds.

Table 3: List of Online Assessments

| Name | Type | Characteristics | Ref. |
| --- | --- | --- | --- |
| Tablet-based assessment tools | NAMs | Early detection of AD; Differentiating MCI and AD; Speech datasets collected from asking question about Daily life; | [123] |
| Telephone Interview for Cognitive Status (TICS) | CoAMs | Usable for both research and clinical practice; Comparable results with MMSE | [124, 125] |
| Rowland Universal Dementia Assessment Scale (RUDAS) | CoAMs | Multicultural, Multi-language | [124] |
| Computerized ADAS-Cog (cADAS Cog) | CoAMs | Higher test-retest reliability [13]than the paper based method | [48] |
| Brain Health Assessment (BHA) [14] | CoAMs | A 10-minute test; Consists of 4-task; Assesses executive functions, processing speed, and visuospatial skills | [126] |
| Computer Assessment of of Mild Cognitive Impairment (CAMCI) | CoAMs | A 20-minute test; Consists of 8 task; Assesses memory impairment, including working and episodic memory, executive functions and visuospatial skills | [127] |
| CogState Brief Battery (CBB) | CoAMs | A 15-minute test; Consists of 4 tasks; Assesses both episodic and working memories, attention, and speed | [128] |
| NIH Toolbox Cognition Battery (NIHTB-CB) | CoAMs | A 31-minute test; Consists of 7 tasks; Assesses episodic memory, language, executive functions, and processing speed | [129] |
| Motor and Cognitive Dual-Tasks (MCDT) | CoAMs | A brain-stress test; Assesses the functioning of the motor-cognitive interface; Distinguishes between several neuro-cognitive disorders Assesses cognitive impairments in individuals with DSL | [130, 131] |
| Word and Sentence Repetition (WSR) | CoAMs | Working memory test for AD; Discover grammatical and extra grammatical errors | [118, 119] |

# 3 AI Assessments for Dementia

Growing user-friendly ML and DL libraries and powerful computational resources graphics processing unit (GPU) and tensor processing Unit (TPU) have motivated AI developers to design, develop and deploy AI assessments for dementia. This section reviews various types of AI assessments (i.e., listed in the alphabetical order), which have been developed to evaluate cognitive and functional impairments or behavioral disorders associated with dementia and its subtypes.

## 3.1 Facial Expression to Develop AI Assessments for Dementia

Dementia, in particular AD, might affects brain's regions such as the amygdala, temporal pole (TP), superior temporal sulcus (STS), and anterior cingulate (AC) and cause difficulties in: 1) Understanding emotions [139]; 2) Decoding facial emotional expressions [140, 141]; 3) Controlling facial muscles [142]; And 4) showing emotional reactions [143, 144, 145].Thus, PwD might have abnormal emotional reactions or problems in understanding emotional reactions. Hence, it has been suggested [143, 144, 141] that developing AI assessments, which evaluate patients facial expressions or facial emotional recognition could be useful to identify patients with dementia and subtypes of dementia. This section, reviews some recent articles related to the above mentioned topics (i.e., Interested readers can refer to the research work done in [145] for getting information about prior articles on facial expression recognition in AD in 2010s and before).

The authors in [143] attempted to develop an AI system to distinguish patients with dementia from cognitively normal individuals by using images of their faces. From 121 patients and 117 cognitively sound participants they collected 484 face images to train five different convolution neural network (CNN) architectures among which Xception [146] model outperformed others. Developing AI systems to assess facial expression recognition (FER) has captured attention. The idea is to assess FER by showing participants some emotional stimuli and then analyze their emotions via the captured facial images/videos. In [141] three intervention sound sources (music, stream, and birdsong as stimuli) were played for 35 older participants with mild dementia while an FER system called FaceReader monitored the evolution of their emotions. FaceReader software scored the intensity of six facial expressions (happiness, surprise, fear, sadness, anger, and disgust) for each frame of the recorded video of the participant listening to each of the stimulus sounds. Furthermore, FaceReader computed the emotional valence (i.e., describes the pleasantness level of emotional stimuli, pleasant =1 and unpleasant = 0) and emotional arousal (i.e., the level of emotional excitement, high = 1 and low =0) of facial expressions. They also compared the results from FaceReader and a subjective emotional evaluation. Throughout their experiments, they found out that fear, sadness, and disgust can significantly vary depending on different stimulus sound source while happiness, surprise, and anger give similar results no matter what sound source is used. Since some stimuli gave different results depending on the intervention date, they recommended using multiple sound sources while using FaceReader.

In [144] video stimuli were played for 61 participants (25 healthy and 36 cognitively impaired) while recording their facial expressions. Their proposed FER system based on MobileNet [147] converts the facial expressions of each participant into a data matrix. The predicted intensity of six emotions (sad, surprise, happy, neutral, angry, and other) for each frame of the recorded video builds such a data matrix. An SVM classifier trained on 46 of these converted data matrices showed a detection accuracy of 73.3% on the remaining 15 participants. Both [141, 144] used a classic object detector called Viola-Jones [148] to detect the faces. Chieti Affective Action Videos (CAAV) database including 360 video stimuli was used in [149] to experimentally control the emotional states of 302 participants. Although their collected data does not include the emotional facial expressions of the participants but rather the scores for both valence and arousal of all the 360 video stimuli. One may use this huge database of video stimuli to record the videos of facial expressions of the participants.

The authors of [150] developed an assessment by adopting multi-modal approaches. The assessment can assess impairments in facial expression associated with MCI. They combined facial expressions with the trajectory motion data obtained from tracking the hand movements and elbow joint distribution in order to monitor early dementia symptoms among the aging deaf signers of British Sign Language (BSL). Recent clinical observations suggest that the signers with dementia may sign differently than cognitively fine signers. These clinical observations indicate that cognitively impaired signers may use a limited number of signs and facial expressions compared to the cognitively healthy signers. In order to distinguish aged signers with dementia from the rest of the signers they proposed a multi-featured deep learning approach which focused on hand movements, body language and facial expressions. They used pose estimation for analyzing the hand, elbow, and body movements. A real-time face analysis model is deployed to identify active and non-active facial expressions. Finally a CNN model is fed with facial and trajectory motion features to identify signers in the early stage of dementia. Having monitored the features, they noticed that trajectory motion ones among signers with MCI create a resemble line rather than the spikes, a trajectory characteristic of healthy signers that tend to make less pauses and static poses while signing. The greater distance between features extracted from different regions of the face is also an indicator that cognitively healthy signers produce more active facial expressions and movements. In their experiments, they segmented videos of 21 healthy and 19 MCI participants into 163 short clips split into the training (80%) and validation (20%) sets. To test the model, they used videos of the five additional healthy and one MCI participants segmented into 24 video clips. They ran their experiments on two CNN models: VGG16[6] and ResNet-50 [152]. Surprisingly the former outperformed the latter by a large margin.

Authors in [153] proposed a multi-modal emotion recognition approach to assess the autobiographical memory deficits associated with AD. To do so, they collected a spontaneous emotion multi-modal database, which encompasses facial expressions and EEG signals of individuals aged between 30 and 60 in order to address the autobiographical memory deficits of patients with AD. They developed an ML-based approach to identify spontaneous reactions from distant and recent memories. They employed an SVM-based classifier and a Gentleboost classifier, which trained using EEG features and facial landmarks as the basis of their proposed approach. For a complete review on multi-modal behavioral approaches for early dementia diagnosis, interested readers can refer to the following article [154].

### 3.2 Eye Movements to Develop AI Assessments for Dementia

Patients with MCI and neurodegenerative disease including dementia have shown impairments in eye movements [155, 156, 14, 157, 158, 159, 160, 161, 162]. Furthermore, it has been stated that PwD might show such symptoms before showing other types of cognitive impairment. Thus, it has been suggested that eye movement patterns should be considered as biological markers to diagnose dementia [159] at its earliest stage or identify patients with neurodegenerative disease including dementia [160, 161, 162]. And clinicians can use Eye tracking technologies [157] (e.gEyeLink 1000 Plus or Tobii Pro) [158] to collect eye movement patterns (e.g., patterns of oculomotor including fixations, smooth pursuit, saccades, and pupil responses [158]). These patterns can be utilized to train AI systems to detect dementia or AD in older people at its earliest stage. The main advantage of such assessments is that they are non-invasive, inexpensive methods. Thus, clinicians can widely use them to detect older adults at risk for dementia [157]. A review [14] of various ET technologies that have been used to develop AI assessments for dementia suggested that ET technologies can be combined with other non-AI assessments to detect cognitive impairment that is associated with dementia and AD. A good example of such a suggestion has been provided in [159]; the authors have developed cognitive assessments using ET technology to detect patients with dementia. Such assessment can be considered as affordable, quick, easy to be administered and accurate [158] cognitive assessments with high sensitivity to detect cognitive impairment.

### 3.3 Neuroimaging-based AI Assessments for Dementia

Neuroimaging techniques such as positron emission tomography (PET) [163, 164, 135, 165, 166], magnetic resonance imaging (MRI) [167], structural MRI (sMRI) and functional MRI (fMRI) [168, 169, 170, 171, 172, 173, 174], electroencephalogram (EEG) [175, 176, 177, 178, 179, 180], magnetoencephalography (MEG) [176, 177, 181, 182, 183, 184], diffusion tensor imaging (DTI) [185, 186] help clinicians do in vivo studies of brain anatomy in individuals to identify patients with dementia [187, 188, 12, 189] or determine patients with specific dementia types such as AD, FDT, VaD [190] with high accuracy [186, 191].

Neuroimaging-based AI assessments have been developed on the basis of ML or DL algorithms [12, 192, 193, 194, 195] using various types of neuroimaging datasets [187, 188] (see Table 4 for the list of neuroimaging datasets). In more detail, DL algorithms such as 2D CNN and 3D CNN have shown high accuracy in differentiating patients with AD from healthy controls[12, 192, 193, 194]. In addition to CNN, other DL algorithms such as Recurrent Neural Networks (RNNs) [195], Stacked Auto-encoder (SAE) [194, 196, 197], Restricted Boltzmann Machine (RBM) and Deep Boltzmann Machine (DBM) have shown reasonable accuracy in classifying patients with AD from healthy controls. Several studies have focused on developing approaches combining DL algorithms (as feature selection methods) with traditional models [12]. Some studies have shown that DL can directly be trained as an end-to-end

---

[6] It is also called VGGNet, a convolution neural network (CNN) model with 16 layers [151]

learning method using neuroimaging data and be employed to diagnose AD [12].

Table 4: Lists of Well-known Neuroimaging Datasets

| Name | Type | Ref. |
| --- | --- | --- |
| ADNI database (Alzheimer's Disease Neuroimaging Initiative (ADNI)) | MRI and PET images | [198] |
| AIBL (Australian Imaging Biomarkers and Lifestyle Study of Aging) dataset | PET | [199] |
| BioFIND dataset (The datasets has been formatted according to international BIDS standards and analyzed freely on the DPUK platform | MEG data and T1 MRI data from individuals with MCI | [200, 182] |
| MIRIAD (Minimal Interval Resonance Imaging in Alzheimer's Disease) databases | Volumetric MRI brain-scans of 46 patients with AD | [201] |
| Open Access Series of Imaging Studies (OASIS) datasets | MRI and PET images | [202] |
| ANMerge dataset | MRI | [203] |
| National Alzheimer's Coordinating Center (NACC) | MRI and PET Images | [204] |

Given the rise of AI applications in neuroimaging and dementia research, ethical considerations related to data privacy and informed consent have become increasingly pertinent. One significant area of concern is the potential for misuse of patient data, particularly given that neuroimaging techniques like PET scans can lead to unexpected privacy issues. For instance, a recent study demonstrated that PET data can be exploited to reconstruct individuals' faces, which clearly presents significant privacy concerns [205]. The sharing and secondary use of neuroimaging data, while invaluable for advancing research, also raises ethical questions. The Belmont Report provides a foundational ethical framework, including respect for persons, beneficence, and justice. As Brakewood and Poldrack discuss in their analysis of the ethics of secondary data usage, these principles can also be applied to the sharing of neuroimaging data [206]. Respect for persons is based on the recognition of the autonomy of individuals and the need to protect those with diminished autonomy. In the context of AI and neuroimaging, this implies obtaining informed consent for data collection and subsequent sharing, see [207, 208]. Patients must be informed about how their data will be used, the potential for face reconstruction from their PET scans, and other potential privacy risks. Ensuring that participants comprehend these risks is also critical, especially considering that dementia patients may have impaired cognitive abilities [209]. The principle of beneficence involves ensuring the well-being of individuals by maximizing possible benefits and minimizing potential harms. The sensitive nature of neuroimaging data necessitates the implementation of stringent measures to safeguard patient privacy and prevent data misuse. Data anonymisation and advanced cybersecurity practices are some of the ways to minimize potential harm. Justice, the third principle, refers to the fair distribution of benefits and burdens. Here, it would imply that the benefits of neuroimaging research (like advancements in diagnosis and treatment of dementia) should be accessible to all, and not just a subset of the population. It also means recognizing and addressing any potential biases in the data that could disadvantage certain groups. Moving forward, ethical guidelines and policies need to be regularly updated and enforced to keep up with advancements in AI and neuroimaging techniques. Involvement of ethicists, legal experts, data scientists, and the patient community in this process is vital for developing comprehensive and practical guidelines. Furthermore, creating a culture of ethics within the scientific community, where researchers are committed not just to advancing knowledge but also to respecting patient rights and welfare, is of paramount importance.

The main benefit of using AI-based neuroimaging techniques for dementia is that they can help the medical community to: 1) Provide an accurate diagnosis, particularly when they suspect that an individual might suffer from MCI or AD; 2) Use detailed images of brain structure or functional activities for better investigation of dementia and AD; 3) Provide supportive evidence that allows them to accurately detect dementia; 4) Diagnose dementia types such as AD at the most initial point (i.e., so that patients can take advantage of both pharmaceutical and non-pharmaceutical treatments, which could help them to slow down the progression of the disease [177]); 5) Use it as non-invasive methods in a sequential manner to provide an accurate diagnosis of dementia. For example, MRI is the first-line method capable of detecting cognitive impairment at its earliest stage. It has been stated that MRI techniques can diagnose some types of dementia with high sensitivity and specificity [210]. While fMRI is useful as a second- or third-line investigation to provide a more accurate diagnosis of dementia when there is a doubt and distinguish different subtypes of dementia, particularly FTD from AD. The main drawback of using AI-based neuroimaging techniques is that they are not cost-effective. Furthermore, current insights are generally based on the specific datasets that are publicly available, leading to issues of selection bias/dataset decay that may limit the generalisability and external validity of findings made from these datasets [191, 211].

## 3.4 Sensor-based AI Assessments for Dementia

One approach to develop AI assessments for dementia is using wearable sensors (e.g., accelerometers) [212], non wearable sensors, assistive sensors, and physiological measurement devices (which can measure individuals' blood pressure), blood oxygen saturation, temperature), smart home and environmental sensors (e.g., infrared motion detectors and magnetic door sensors) [15, 213, 214, 215, 216, 217, 218, 219, 162]. This section reviews studies, which have focused on developing sensor-based AI assessments.

### 3.4.1 Home and Environmental Sensors to Develop AI Assessments for Dementia

Smart home and environmental sensors are the basis of the development of interactive smart home systems with the great potential of delivering assistive technology to PwD and the older adults at the risk of dementia [220, 221, 222]. The main benefits of assistive technology for PwD is that they can live independently at home while at the same time avoiding institutionalizing [223, 224, 225, 226]. Other benefits of developing assistive technologies for older adults and PwD are: 1) to increase their independence, memory function, quality of life (QoL), safety and security [227, 228, 229, 230], 2) improve treatment routines by using medication dispensers or smartphone applications, while providing reminders and notifications [231, 232], 3) address dementia-related problems, such as incontinence, by developing wetness-sensors and urine detection devices [233, 234], or recognize activities of daily living (ADL)[235, 236], 4) offer support to elders regarding time and orientation [237, 238] (especially using a combined approach that use wearable sensors such as accelerometers and wireless heart rate monitors or audio and/or audio-video prompts), 5) assist people with memory impairment through prompts on a daily basis [239, 240, 241].

Furthermore, more complex solutions, the so-called "smart homes", include several sensors to monitor users' ability to cope with ADL or prevent significant incidents [242] and observe and evaluate PwD activities [243]. Additionally, more severe levels of dementia have been explored using sensors to recognize and assess not only physical activity but also vital signs by providing alarms in case of emergencies [244]. Other systems help people with cognitive deficits to remember ADLs (e.g., instructions on how to prepare a hot meal etc.) [224]. Moreover, sensors have been attached to electrical appliances (a cooker, refrigerator, TV, etc.) to detect events caused by disease or possible accidents and identify behavioral trends [225]. A recent study evaluated the possibility of using unobtrusively collected activity-aware smart home behavior data to detect the multimodal symptoms of AD from older adults for over two years, containing behavioral features, mobility, cognitive and mood evaluation [214]. Another one developed the night-time wandering detection and diversion system to assist caregivers and PwD that are at risk of wandering at night, using sensors, smart bulbs, pressure mats, speakers and prerecorded audio prompts [245].

Recently a design and deployment of a smart home in a box (SHIB) approach to monitoring PwD wellbeing within a care home was tested, where sensors were installed to detect ADLs [246]. More recently, researchers developed an in-home assessment of ADLs over many months to years in order to predict cognitive decline in community-dwelling older adults wearing fitness trackers assessing daily sleep and physical activity patterns, a sensor-instrumented pillbox on a daily basis [247]. On the other hand, the CART platform [248] comprising ambient technology, wearables, and other sensors, was deployed in participants' homes to provide continuous, long-term and ecologically valid data to 232 elders. Multiple measurements of wellness such as cognition, physical mobility, sleep and level of social engagement, showing that CART initiative resulted in an unobtrusive digital health-enabled system that allows data capture over extended periods and monitors health remotely. A recent study presented a conceptual computational framework for the modeling of ADLs of PwD and their progression through different stages demonstrating that it is feasible to effectively estimate and predict common errors and behaviors in the execution of ADLs under specific assessment tests [249].

Another holistic approach investigated the long-term effects of assistive technology combined with tailored non pharmacological interventions for PwD and received the system installed at home for 4 to 12 months. The participants, who received the sensor-based system, have shown improvement in domains such as sleep quality and daily activity, as measured by the multi-sensor system compared to control groups who did not receive anything. Detection of sleep patterns, physical activity, and ADLs, based on the collected sensor data and analytics, was available at all times through comprehensive data visualization solutions [243, 250]. The above mentioned studies have proved the usefulness of smart home systems in supporting clinicians to reliably drive and evaluate clinical interventions toward quality of life improvement of people with cognitive impairments across AD.

### 3.4.2 Wearable and Optical Sensor to Assess Gait Patterns

Studies [251, 252, 253, 254, 255, 256, 257, 255, 252, 258, 253, 254, 259] suggested that individual's gait pattern, which is evaluated on the basis of walking speed, cadence (i.e., number of steps per unit of time), walking base width (i.e., measured from midpoint to midpoint of both heels), step length (i.e., measured from the point of foot contact to the point of contralateral foot contact) and stride length (i.e., linear distance covered by one gait cycle), might be impaired because of cognitive impairment [251] associated with dementia. Furthermore, these studies have reported that there is a correlation between gait patterns (e.g., gait behavior, gait cycle, gait speed, and walking speed) [255], and cognitive impairment stages [256, 259] associated with dementia. Thus, clinicians might collect gait patterns including decreasing in stride length and walking speed and increasing in support phase [257] from PwD [257, 252, 258, 253, 254] using wearable sensors and optical sensors. Gait patterns can be used to develop AI assessments that can aid clinicians to detect gait disorders, which are more common symptoms of dementia. Since, gait disorders could be related to the severity of cognitive decline [257], thus gait pattern-based AI assessments can be also used to differentiate patients with different types of dementia including [255], AD and DLB [252, 253, 254].

Optical sensors such as video cameras enable real-time gait analysis [260]. The goal is to find the most important patterns for determining the presence of neurodegenerative disease including dementia from the earliest signs. To do this, the machine must be able to recognize fine grained movements typical of the disease, such as tremor and freezing gait in Parkinson's. In this particular direction the work in [261] suggests performing pose estimation on video recordings of people with dementia and control subjects. The body joints coordinates in time extracted by the pose estimation algorithm are then smoothed by using a Kalman filter and, sequentially, the steps are recognized and segmented. For each step several features have been extracted such as velocity, displacement, acceleration as well as sigma-lognormal features from the Kinematic Theory of Rapid Human Movements (KTRHMs) [262]. The accuracy reached in binary classification is as high as 0.95. The evolution of this work is presented in [263] where a deep neural network (DNN) composed of two parallel branches, one slow and one fast, looks at the data at two different speeds. Similarly to a slide viewed by a microscope at two different resolutions, the Multi-Speed Transformer (MST) looks fast paced patterns, such as the presence of a false start, also known as freezing gait, as well as fine grained movements, such as tremors. The first branch of the architecture takes as input the raw time-series of body joints coordinates, performs a 1D convolution with strides of 1 and dilation rate of 2 followed by a multiheaded attention with positional encoding. While the second, fast branch, performs a 1D convolution on the input with strides of 3 and dilation rate of 2 followed by multi headed attention with positional encoding. The embeddings are concatenated and a standard feed forward neural network completes the classification or regression. The accuracy, on the exact same conditions is as high as 0.97.

Clinicians can evaluate gait patterns to detect patients that might develop dementia, different dementia subtypes. Using gait patterns, monitoring severity of dementia would be as well possible[264]. Interested readers can refer to a systematic review of gait-based assessment methods, which have been done in [265] as well as in [260].

### 3.4.3 Wearable Sensors to Assess Sleep Disorders Associated with Dementia

Sleep disorders including insomnia, changes in wake-sleep rhythm are common disorders among patients with dementia in particular patients with AD and amnestic mild cognitive impairments [266]. Patients with AD and dementia experience sleep disturbances including changes in daytime sleep, slow wave sleep and rapid eye movement sleep [267]. Thus, it has been suggested that wearable sensors (e.g., WHOOP) can collect sleep cycles and monitor sleep routines in older adults.

### 3.4.4 Digital Pen to Develop AI Assessments for Dementia

Prominent symptoms of people affected by dementia are neuro-muscular diseases. The peripheral nervous system is affected by neuro-muscular diseases, because of the presence of sick or dead neurons. Thus, when messages sent by the nerve cells try to control these muscles and perform movement, the signal is corrupted resulting in an uncontrolled movement or the total lack of movement.

Handwriting movements are the result of a complex network mainly composed of cognitive, kinesthetic, and perceptual motor abilities [268, 269]. Handwriting patterns have been found to be prominent features for developing AI for dementia assessment as well as lots of other neurodegenerative diseases. This is because the subject's brain performing handwriting undergoes synergic cooperation of various areas of the brain, such as basal ganglia, cerebral cortex and cerebellum [270, 271]. Only recently, it has been investigated the use of handwriting performed with a digital pen and a tablet in order to capture handwriting signals both online (time-series of x,y coordinates usually integrated with pen status, if pen touches the surface or it is on air and azimuth) or offline (color or b/w image of a handwriting signal).

Often, people affected by dementia undergo a clinical investigation only after the disease has had a visible onset. For this reason, it is important to focus on non-invasive biomarkers capable of finding the early sign of the disease. Thus, identifying accurate biomarkers allowing differential diagnosis, prompting clinicians to investigate and start therapy, as well as analyzing the response to the therapy, are the primary goals of the research on dementia [272]. In this regard, handwriting can fill a prominent role; in fact, handwriting is a complex activity enabling the synergistic interaction of various parts of the brain whose decline is reflected in clear handwriting patterns that could be captured in early stages allowing, thus, a prompt activation of the medical diagnosis and thus treatment. In this regard, it has been observed that the abnormal reduction of the size in handwriting, also known as Micrographia, can be easily found in patients with Parkinson's disease, while Dysgraphia, which is the progressive degeneracy as well as confusion in handwriting, can be found in patients with Alzheimer disease as well as dementia [269, 268]. Over the years various computer-aided diagnosis systems have been developed with the aim of supporting the early diagnosis of dementia and in general neurodegenerative disease as well as a tool for analyzing the progression of the disease. The majority of works focus on creating a computerized predictive model capable of finding subtle changes in handwriting. These models are then used to perform inference and thus find the presence of the disease with a certain level of accuracy. The acquisition tool is, usually, a tablet with an electronic pen. The electronic pen is capable of capturing temporal and spatial signals. These signals are then converted into features or are processed as-is. In facts literature can be divided into handcrafted features, which also introduces the use of physical models developed to reconstruct wrist and hand movements, and into deep learning models which use deep neural networks that work on raw time-series or convert the signals into images and then use standard convolutional neural networks to perform classification on images. Despite the results achieved by the community, there is also the additional problem that the existing datasets are not homogeneous. This results in datasets with different handwriting tasks, not harmonized and not interchangeable, which in most cases, also provided different results. The only way to overcome this problem is to use a shared acquisition protocol. Authors in [273] and in [274] have used a digitized version of standard handwriting tests accepted and tested in the neurological and psychological community, which can be evaluated with standard scales such as Milan Overall Dementia Assessment (MODA) [275] or MoCA, and thus can be used as ground truth for evaluation.

Starting from x,y coordinates as well as button status (the in air movement), pressure, azimuth and altitude, usually the majority of related works extracted the displacement over the two coordinates, the velocity, the acceleration, the Jerk, the number of changes in velocity and in acceleration as features to be fed into a classifier model [276]. In addition to these features, authors also used different velocity-based features dictated by neurodegenerative disease's motor deficits, such as bradykinesia (slowness of movements), micrographia (reduction in size of writing), akinesia (impairment of voluntary movements), as well as muscular rigidity and tremors. These additional models are the Maxwell-Boltzmann distribution, which is used to extract parameters that model the velocity profile and use them as features for the classification model, and the Sigma-lognormal model from the Kinematic Theory of Rapid Human Movements [277, 276, 278]. This theory is an instrument to analyze handwriting movements as a statistical process that leverages on neuromuscular parameters of the human body and brain [262]. The model is capable of reconstructing the final result of movements, in terms of acceleration and velocity profiles. This theory is defined by using two main components: the agonist neuromuscular system, acting in the same direction of the movement, and the antagonist neuromuscular system acting in the opposite direction. This theory is based on the intuition that movements are the combination of primitives, also named strokes, having a lognormal acceleration and velocity profile. Thus, a handwriting trace is made of several strokes connecting a sequence of points. The parameters of this sigma-lognormal model are used as features. Depending on the task and dataset, accuracies of dementia assessment through handcrafted features go well above the 90%, but also it is not uncommon that different features have different importance depending on the task.

In this section, we refer to deep learning algorithms by differentiating algorithms that make use of images as input from algorithms that make use of raw coordinate time series. The most used DL algorithms that take in input images are CNNs. In this setting, handwriting and drawings are converted into images, usually, on-air movements and on-surface movements have different colors. For example authors in [279] used blue for on-surface movements and red for in-air movements. This time series of coordinates are plotted and the resulting image is used to train or validate the CNNs. Several CNNs architectures have been used, the most famous are ResNet, Inception V3, NASNet, Inception ResNet v2 and VGG19 [279, 280]. Another way of using Deep Learning for dementia recognition through handwriting is by using recurrent neural networks RNNs in particular Long-Short Term Memory Networks LSTM [281] and transformers [282].

Datasets for Developing Handwriting-based AI Assessments There are different datasets (PaHaw [283], NewHandPD [284], ParkinsonHW [285], ISUNIBA [286], EMOTHAW [287] and Darwin [288]) for dementia and neurodegenerative disease assessment through handwriting. The PaHaW Database [283], which includes numerous handwriting samples, was compiled using samples from 37 Parkinson's disease patients (19 males/18 females) and 38 controls of comparable gender and age (20 males/18 females). This database was created in partnership with the Movement Disorders Center at Masaryk University's First Department of Neurology and St. Anne's University Hospital in Brno, Czech Republic.

The NewHandPD dataset [284] uses signals from spirals and meanders and is composed of 35 individuals, with 14 patients (10 males and 4 females) and 21 healthy controls (11 males and 10 females). Each participant was instructed to use a smart pen and begin filling out the form from the inside and working outwards.

The ParkinsonHW dataset [285] consists of 62 individuals with Parkinson's disease and 15 healthy individuals. The dataset includes three types of handwriting recordings for all participants: the Static Spiral Test (SST), Dynamic Spiral Test (DST), and Stability Test on Some Points (STCP). The ISUNIBA [286] was created by collecting data from 29 individuals who likely have Alzheimer's disease and 12 healthy control participants. These individuals were instructed to write the Italian word for "mother" over several writing sessions. This word was chosen because it is one of the earliest words learned in life and often one of the last words spoken before passing away.

The EMOTHAW [287] (Emotion Recognition from Handwriting and Drawing) dataset was developed as a tool for recognizing emotional states. This dataset utilizes 129 participants' handwritten texts in Italian language and drawings to detect emotional disorders and negative emotions, such as anxiety, depression, and stress.

The DARWIN [288] (Diagnosis AlzheimeR WIth haNdwriting) dataset comprises handwriting samples collected from 174 participants, including both patients and healthy controls. Of these, 89 participants were classified as Patients (P) and 85 as Healthy (H).

Almost all the reviewed datasets contain raw signals, such as time-series of pen coordinates, images or already synthesized features like position, button status, pressure, azimuth, altitude, displacement, velocity and acceleration. Unfortunately, usually, these datasets do not share the same tasks and it is thus difficult to perform cross-datasets testing.

### 3.5 Voice-based AI Assessments for Dementia

Voice-based AI assessments can automatically detect speech and language disorders associated with dementia [289, 290, 291, 292, 293, 294, 295, 296, 297, 298, 299, 300, 301, 302, 303, 304, 305, 306, 307, 308, 309, 310, 311, 312, 313, 314, 315, 316, 317]. Thus, these systems can help clinicians identify older adults with dementia from their voices and languages, which are obtained by engaging participants to do an ordinary conversational task or complete a series of language tasks [318, 319, 320, 321, 322, 123, 323]. Table 5 lists different types of voice-based AI assessments that have been developed to detect patients with dementia (in particular AD) or MCI from their speeches [324] or their spontaneous speeches [325, 326, 327, 328, 329, 86, 330, 331].

Generally, to develop such systems using traditional ML algorithms the following stages could be followed: 1) Collecting vocal and language datasets or getting access to available speech and language datasets; 2) Pre-processing data; 3) Using feature engineering methods including feature extraction and feature selection to select informative features and feature fusion techniques; 4) Choosing an ML algorithm, training it, evaluating its performance and validating its results. The rest of this section has focused on: 1) Describing speech and language disorders associated with dementia; 2) Reviewing vocal and verbal tasks; 3) Listing benchmark speech and language datasets [325, 332, 323, 333, 334, 335, 315]; 4) Highlighting speech and language features; 5) Referring to developed ML and DL algorithms [299, 325, 330, 336], which have been the basis of voice-based AI systems [337, 338, 87, 339,

321, 340, 341].

### 3.5.1 Collecting Patients' Voices

One essential step to develop voice-based AI assessments is to collect vocal data or get access to collected vocal datasets. Several vocal and verbal language tasks have been proposed to collect patients' voices. In the following, we have described some language tasks:

1. Alpha-span Task evaluates verbal–numerical working memory impairment in patients with dementia [344]. During this task, examiners present a series of words to subjects and ask them to recall words in alphabetical order.

Table 5: Voice-based AI Assessments

| Name | Type | Ref. |
| --- | --- | --- |
| Detect patients with AD | Patients' spontaneous speech | [293] |
| Identify patients with MCI, pre-dementia and early dementia | Patients' spontaneous speech | [327, 328, 329, 86, 330, 331] |
| Detect speech rhythm alterations associated with AD | Patients' speech in Spanish | [342] |
| Distinguish patients with MCI from healthy controls | Patients' speech | [329] |
| Distinguish patients with pre-dementia from those with AD | Patients' speech | [86] |
| Identify AD | Speech | [330, 331] |
| Develop AD risk assessment | Speech | [292] |
| Detect MCI | Spontaneous speech | [296] |
| Detect AD | Spontaneous speech | [293] |
| Distinguish mild AD and MCI | Patients' speech | [309] |
| Distinguish patients with AD from healthy controls | Patients' speech | [308] |
| Detect AD | Patients' speech and transcriptions | [301, 297] |
| Assess AD | Patients' speech and transcriptions | [123] |

2. Auditory Sentence-to-picture Matching evaluates sentence comprehension [343] and encompasses nine different types of sentences with varying levels of syntactic complexity [345].

3. Digit Span Task (DST) evaluates verbal attention and working memory in participants [346]. During the task, a subject is asked to recall the forward order (or Digit Span Forwards (DSF)) or backward order (Digit Span Backwards (DSB)) of a string of digits that has been presented by an examiner [155]. The DST in particular the DSB can be used to identify patients with major cognitive impairment [133].

4. Hopkins Verbal Learning Test (HVLT) examines the functionality of the verbal memory, which is one of the preliminary signs found in amnestic Mild Cognitive Impairment (aMCI) and other types of dementia [76, 77, 78]. The HVLT encompasses a 12-item word list composed of four groups of words corresponding to three semantic categories. The subject is asked to listen carefully as the examiner reads the word list and attempts to memorize the words. The main advantage of HVLT is that it is a quick test. Another benefit of HVLT is its usefulness in evaluating cognitive impairments at frequent intervals [347].

5. Picture Description Task (PDT) evaluates the semantic knowledge in subjects [291] and assesses their structural language skills [348]. During the task, clinicians can use a picture mostly "the Cookie Theft", from the Boston Diagnostic Aphasia Exam, or other pictures such as "a man changing a lightbulb" and "a kitten in a tree" and ask participants to describe what they see in the picture [349]. It could be used to provide monologue speech from participants [349].

6. Sentence Repetition Task (SRT) is a test for verbal short-term memory and can be used to detect language difficulties. Since, patients with dementia have a short-term memory problem that causes them to repeat words (i.e., word repetition in people with dementia ). The SRT test is useful to detect an abnormality in repeating words associated with dementia.

7. Verbal Fluency Test (VFT) measures 1) phonemic fluency (Letter Fluency) or 2) semantic fluency, semantic verbal fluency (SVF) or category fluency. For evaluating phonemic fluency, clinicians use the Letter Fluency Task (LFT) and ask participants to create as many words as possible starting with a specified letter (e.g., M) [350, 351] to test verbal functioning of participants. COWAT-FAS is a similar task to Letter Fluency Task [352]. For measuring semantic fluency, participants are asked to complete *Category Fluency Task (CFT)* or *Semantic Fluency Task* or *Semantic Verbal Fluency (SVF) Task* by creating as many words as possible in a specified category (e.g., food) during 1 minute[353], [354] It has been shown that VFT as a short cognitive test to assess verbal memory impairment is helpful to discriminate between healthy aging and very mild dementia of the Alzheimer type[7] [351].

8. Semantic Association Test (SAT) evaluates disorders in verbal and visual semantic processing in older adults [355].

9. Object Picture Naming (OBN)/ Picture Naming Test/the Boston Naming Test (BNT) has been widely used in determining impaired picture naming and cognitive decline [356] in PwAD or MCI [357]. Impaired picture naming is a result of semantic network degradation. It assesses language performance in participants with dementia and AD. It can detect compromised lexical retrieval abilities and aphasia via visual altercation naming [358]. It can be used as part of a cognitive assessment method called Consortium to Establish a Registry for Alzheimer's Disease (CERAD), the performance of BNT can be affected by educational level, sex, race age, and living environment of the elderly [359, 360]. Note that this test is also a part of MoCA [343].

10. California Verbal Learning Test (CVLT) assesses memory (verbal memory) and language capacities including immediate recall, free and cued recall over short and long delays, and Yes/No Recognition [361, 362, 369]. Different variations of CVLT, including CVLT-II and CVLT-III have been proposed [83].

11. Story Recall Test (SRT) or delayed story recall evaluates episodic, semantic and verbal memory [363]. The SRT (i.e., during the story recall test, participants are shown a short passage with one of the following options 1) My Grandfather, 2) Rainbow or 3) Limpy, which are three well-known passages to assess memory capacity of participants) can assess impairment in episodic and semantic memory, and also global cognition [363, 364]. It would be a suitable test for subjects with a high level of education rather than a low level of schooling [365] and to predict progression to dementia in patients with MCI [366], the Automatic Story Recall Task (ASRT) has been developed as a fully automated SRT task that can be remotely run [84].

12. Alzheimer's Disease Assessment Scale– WORD RECALL Test (WRT) measures the language ability of individuals by giving them three trials to learn 10 words that can be considered as high-frequency, high-imagery nouns and have been printed in block letters on white cards. The subject's score can be calculated as the mean number of words not recalled on three trials [367].

13. Alzheimer's Disease Assessment Scale–word recognition task measures memory impairment. During this task subjects are given a list of 12 words to learn and then examiners show them a series of words and ask subjects if she/he can remember them [367].

14. Detection Test for Language impairments in Adults and the Aged (DTLA) is a quick language task that has been designed to assess language problems associated with neurodegenerative disorders [343].

15. Arizona Battery for Communication Disorders of Dementia (ABCD) encompasses 14 subtests to evaluate linguistic expression, linguistic expression and comprehension, verbal episodic memory, visuospatial construction, and mental status [87, 88].

16. Word and Sentence Repetition (WSR) measures working memory impairment in Alzheimer's patients. Clinicians

---

[7] The Verbal Fluency Test for Dementia Screening

use this task to discover some dementia's language markers such as grammatical and extra grammatical errors [118, 119].

17. Spontaneous Written Sentence Production (SWSP) is a realistic assessment of various hierarchical levels of language organization, including morpho-syntactic and lexico-semantic levels [304]. Some of the vocal and verbal tests mentioned above are a part of comprehensive cognitive assessments, while some can be considered stand-alone neuropsychological assessments[352]. They can also be used to collect speech and language samples from patients and healthy controls [363, 363, 370].

Tasks such as MBT, VFT, SVFT, PDT, SRT and countdown tasks are commonly used to obtain speech data from individuals [316] that can be used by clinicians to assess speech and language disorders [343] associated with dementia.

### 3.5.2 Getting Access to Collected Individuals' Voices

Since, a comprehensive report about speech and language datasets has been provided in [371] this part mainly has focused on describing two popular and publicly available vocal datasets. Some other vocal datasets have been listed in Table 6. DementiaBank(DB) is a dataset[372, 301, 308, 373] collected as a part of a project named TalkBank, which has been done as a part of the Alzheimer Research Program at the University of Pittsburgh. The DB contains eight corpora in English, one in German, two corpora in Mandarin, one in Spanish, and one in Taiwanese. The main corpus in the DB is the Pitt corpus [333, 334, 335, 374, 314], which has been collected longitudinally, between 1983 and 1988, every year from around 200 patients with AD and 100 healthy controls. The Pitt corpus contains several sub corpora, generated according to neuropsychological tasks performed by the participants: 1) the Cookie Theft of the PDT (note that the participants' speech related to this task have been transcribed), 2) the word 3) LFT, 4) the SRT, and 5) the sentence construction task. Note some tasks performed either by healthy and dementia participants or only by dementia participants. Several studies [333, 375, 376, 377, 378] have used Pitt corpus to develop systems to detect dementia [333], to identify dementia [375], to diagnose AD [376], to detect cognitive impairment [377].

Alzheimer's Dementia Recognition through Spontaneous Speech or ADReSS Challenge Dataset has [325, 332, 323, 320, 293] been developed by modifying the Pitt corpus. It includes audio and text files for each subject and has a balanced distribution in terms of classes, age, and gender. In more detail, the ADReSS Challenge dataset includes data from 82 AD and 82 non-AD participants, of which 54 AD and 54 non-AD participants are included in the train set[8].

### 3.5.3 Challenges in Collecting Vocal Datasets

Collecting high quality (i.e., accurate, valid, reliable and consistent, complete, comprehensive and granular and unique) vocal samples is a complex procedure due to stringent regulations that should be followed by data collectors [385] and lack of a standardized procedure. Collecting vocal data from patients with dementia or MCI is challenging due to lack of principles for 1) having effective communication with patients [386], 2) training people to correctly collect the data, 3) assuring the quality of recording media and designed experiment behind data collection process that might be led to collect biased, incomplete, invalid datasets [385]. Another issue is rising from the absence of multicultural vocal and verbal tasks that are the basis of experimental designs. Moreover, those tasks are not suitable to assess individuals with intellectual disabilities or persons with hearing loss. And there is no standard approach or agreed-upon assessment methods to collect data from these types of individuals [137]. Furthermore, verbal and vocal tasks such as picture description tasks are highly sensitive to restrictions in sensory input. It has been stated that a person with a slight visual problem has lower performance in completing these tasks [138].

### 3.5.4 Speech and Language Features

Different studies have suggested various types of features to develop voice-based AI assessment for dementia.

*Speech Features* can be provided from vocal samples, collected using vocal tasks [316] including PDT. Articles use various terminologies such as vocal features, acoustic features or audio features in order to refer to speech features. Reviewing various articles, we have categorized speech features into following groups:

1. Voice Activity-related Features (VAF) could be used to measure the temporal characteristics of speech [387] and include silence-related features (SiF), hesitation-related features (HF). Note that tasks such as VFTs can be used to provide vocal samples that are suitable for extracting VAFs in particular silence rate (SR), response length (RL) and the ratio of silence to spoken duration. The VAFs can be used to detect the presence and severity of cognitive impairment caused by dementia as well as to differentiate between patients with AD and MCI [388]. The main motivation behind extracting VAFs is that the temporal characteristics of spontaneous speech including speech tempo, number of pauses in speech, and their length could be considered as vocal markers of the early stage of AD [387].

---

[8] To get more information about the ADReSS Challenge dataset, readers can refer to Alzheimer's Dementia Recognition through Spontaneous Speech The ADReSS Challenge.

The VAFs include features such as duration of speech, speech or speaking rate (SR) or articulation rate (AR) [329, 389], pause rate (PR)[306], number of pauses (No.P) or pause frequency, total length of pauses (TLP), average length of pauses (ALP), duration of pause (per utterance)[316, 306, 389], response time (RT), response length (RL), silence rate (SR), silence-to-utterance ratio, long silence ratio (LSR), average silence count (ASC), continuous speech rate (CSR), average continuous word count (ACWC), hesitation ratio (HR) [390, 321], rhythm (i.e.,rate of syllable production during speech) [342, 390, 306].

2. DL-based Features (DF) could be used to develop end-to-end speech assessments. The DF include I-vectors and X-vectors [308], VGGish-based features, YAMNet-based features [391], BERT-based features [391], CRNN-based features [324].

Table 6: Lists of Vocal Datasets

| Dataset Name | Characteristics | Ref. |
| --- | --- | --- |
| Audio Datasets of Framingham Heart Study | Longitudinal audio files from 1264 subjects, 483 were of participants with normal cognition (NC), 451 recordings were of participants with mild cognitive impairment (MCI), and 330 were of participants with dementia (DE); English | [379] |
| Carolina Conversations Collections | Conversations between adults older than 60 years and young interviewers; 125 speech older speakers without any impairments and 125 speech from speakers with dementia; English | [380] |
| Audio Datasets of the Dem@Care Project | Three different audio datasets that are referred to as 1) Ds3 (i.e., The audio files provided from microphone); 2) Ds5 (i.e., The audio files from from microphone from Dem@Care short @Lab protocol); 3) Ds7 (ie., the audio files from microphone from Dem@Care long @Lab protocol); Greece | [381] |
| Japanese Elder's Language Index Corpus (JELiCo) | Corpus was collected from 22 individuals aged 74 to 86 years JELiCo; Japanese; Multi-educational test | [382] |
| Dementia Blog Corpus (DBC) | Corpus that has been built of several thousand blog posts | [383] |
| Wisconsin Longitudinal Study (WLS) | Longitudinal dataset (six times between 1957 and 2011); 10,317 graduates, sibling and spouses of graduates; the "Cookie Theft" task; English | [384] |

3. openSMILE's Features include three subsets of features: *emobase*, *eGeMAPS*, and *ComParE* features [392, 393, 394, 395, 308, 396, 397]. The *emobase* set includes features such as the mel-frequency cepstral coefficients (MFCC) and fundamental frequency (F0) [397]. The *eGeMAPS* feature set encompasses affective prosodic features [398] such as the F0 semitone, loudness, spectral flux, MFCC, jitter, shimmer [329, 311], F1, F2, F3, alpha ratio, Hammarberg index, and slope V0 features, as well as their most common statistical functional [397, 399, 323]. *ComParE* features [395, 308, 396, 397, 392, 393] include emotional features from individuals' speech [394].

4. PRAAT's Features include a set of vocal features, i.e., prosodic features such as intonation, stress, and rhythm [400]. The PRAAT's features can be extracted using the open software tool PRAAT or Praat vocal toolkit [318, 336]. Some studies have shown that the PRATT tool is useful to extract *F0- based features* (e.g., mean F0) [400], rhythm [400], and *voice quality (VQ) features* (i.e., which can be considered as markers for MCI [329]) include jitter-based features, shimmer-based features and harmonicity-based features [401, 400].

5. Prosodic Features (PF) include fundamental frequency and formants, and intensity features and mel-frequency cepstral coefficients [309], stress, intonation, and emotion [321]. The PF can be used to identify the effect and articulation in speech [402]. Studies suggest that patients with AD have problems in production, repetition, and comprehension of emotional prosody [403], other studies have shown that patients with AD and other dementias have prosodic impairment [404]. Thus, it has been stated that AI assessments, which are developed by training ML classifiers (such as SVM) using PF [402] can identify patients at the late-stage of Alzheimer's.

6. Librosa Features (LF) include features, which can be extracted using the Librosa Python Package [405, 298]. The LF can encompass Librosa spectral features [406] including mel-frequency cepstral coefficients (MFCCs) and Log-Mel spectrograms, their delta, and delta-delta.

7. Temporal Disfluency Features (TDF) includes features such as duration of speech, pauses, prolongation, filler words (e.g., um, uh, hmm [349]), interposed words, question words.

*Language Features* Language features or linguistic features, which include lexical [294, 407, 408, 409], syntactic [407, 410], semantic, and pragmatic [411], psycholinguistic features, can be extracted from patients' transcripts or computer-based or hand written tasks. Studies showed that different linguistic features are associated with different types of language deficits in PwD [412]). These linguistic features include incoherent speech, tangentiality [289] and grammatical error [305, 387], lexical retrieval difficulties, auditory comprehension difficulties, grammatical and spelling failures [413, 305, 387, 290], which have been collected from patients' speech transcript [414], can be markers of language impairment in older adults [414]. Studies have shown that Python libraries including Natural Language Toolkit (NLTK) [415], Gensim [416] and spaCy [417] can be employed to extract linguistic features, which can be used to train ML and DL algorithms to develop language-based assessments [321, 316, 290].

Language features can be classified into different groups such as 1. Lexical Features (LF)[407, 408, 368, 409]; 2. Syntactic Features (SF) [331]; 3. Pre-trained Language Model Features; 4. Semantic Features; 5. Pragmatic Features; 6. Sentiment Features. In the following, each category of features has been described in detail.

1. Lexical Features (LF) [407, 408, 368, 409], encompasses various types of features such as Lexical Repetition (LRe) [407], which can be calculated globally and locally, including ratio of of repetitions (RR) and ratio of self-corrections (RC) or changes, the frequencies of repetitions [337, 339, 418, 419]; Lexical Specificity (LS) [407], which include the number of utterances per individuals (No.U)[9] [341], the number of verbs per utterance (No.V/U) [420], the number of word per utterance (No.W/U) (it can be considered as verbosity [337]), and utterance rate (UR) ,mean length of utterances (MLU) [321, 330, 341, 421], GoAhead Utterances (GAU) [422], the number of function words (No.fW), the number of unique words (No.uW), the number of word (No.W), character length (CL), the number of sentences (No.S), repetitions, revisions, morphemes, trailing off indicator, the number of incomplete words per utterance (No. iW/U) , the number of filler words per utterance (No. FiW/U) [341]; Hesitation and Puzzlement Features [321] includes question ratio (QR), filler ratio (FR), incomplete sentence ratio (ISR); Lexical Richness or Vocabulary Richness [10] [292, 337] includes features such as type token ratio (TTR), root type-token ratio (RTTR), moving average TTR (MATTR) [321, 330, 294, 323], Brunet's Index (W) [294, 420, 408], Honore's statistic (R) [382, 408, 294, 420, 408, 422]. In addition to the above mentioned features, studies have suggested that other types of features including ratio of various parts of speech or Part-of-speech tags (POS) or POS tagging or grammatical tagging which includes a list of words with their particular part of speech (e.g., nouns, pronoun, adjective, verb, conjunction, preposition, and interjection) [418, 419, 306, 423, 382, 408], number of fluency features (No.FF) [339, 418, 419], potential vocabulary size (PVS)[11], which is a modified version of the TTR, is the estimated vocabulary size [382], and vocabulary level (VL) [382].

2. Syntactic Features (SF) [331] include noun rate (NR), pronoun rate (PR), verb rate (VR), adjective rate (AR), clause-like semantic unit rate, coordinated sentences rate (CSR), subordinated sentences rate (SSR), reduced sentences rate (RSR), number of predicates, average number of predicates, dependency distance, number of dependencies, average dependencies per sentence, production rules [294, 341, 422]. Another important category of the SF includes Syntactic Complexity (SC) or Grammatical Syntactic Complexity(GSC) [424, 292, 407]. It encompasses features Dependency Distance (DepD) [382] and Mean Dependency Distance (MDD) [420] and Yngve

---

[9] The total number of utterances per individual can be considered as a feature to detect language impairment. Each utterance is identified to start from the beginning of verbal communication to the next verbal pause length, such as punctuation [341]

[10] It can be measured by the number of unique words [337] and can be used as markers of probable DAT [294]

[11] The PVS score can be calculated using the extrapolation of the number of types in their narrative samples, more information about calculating the PVS score can be found in [382]

Score [382]; Grammatical Constituents [378, 292]; Psycho-linguistics [292, 337]; Information Content [292]; N-Gram Features [373, 376].

3. Verbal Disfluency Features encompass features such as *Intelligibility Features (IF)* [425], *Lexico-syntactic Diversity (LSD)* and *Word and Utterance Rate (WUR)*. It includes features such as total Utterances errors (TUE), total word errors (TWE), unintelligible word (UW), unintelligible utterances (UU), words/TWE, utterances/TUE.

4. Semantic Features can be used to measure semantic decline in PwD. It has been reported that PwD cannot easily retrieve semantic knowledge, reflecting a semantic decline in their language [316]. Furthermore, the loss of the semantic cohesion is another language sign in PwD and PwAD [426]. Different articles proposed that extracting semantic features from the textual files of patients with dementia are useful to detect semantic decline, incoherent speech, tangentiality [289]

Various ML and DL (see Table 7) that have been the basis of AI systems with exceptional achievements in detecting patients with dementia using their speech and language [299, 325, 12, 314]. Among ML algorithms, the SVM-based classifier is a popular algorithm to develop these AI systems; while among DL algorithms, the CNN-based classifier is a trendy DL algorithm.

Table 7: List of ML and DL algorithms have been used to develop Voice-based AI assessments

| Name of Algorithm | Python Libraries of Algorithm | Ref. |
| --- | --- | --- |
| k-Nearest Neighbor (kNN) | Scikit-learn | [422, 420, 378, 333] |
| Support Vector Machine (SVM) | Scikit-learn | [321, 316, 290, 299, 341, 427, 86, 422, 333, 306, 123, 428, 326, 429, 430, 378] |
| Decision Trees (DTs) | Scikit-learn | [422, 431, 290] |
| Random Forest (RF) | Scikit-learn | [432, 330, 420, 336, 306, 326, 378, 293] |
| Logistic Regression (LR) | Scikit-learn | [433, 336, 309, 429, 337] |
| Naive Bayes (NB) | Scikit-learn | [306, 337] |
| Gradient boosting (XGB) | Scikit-learn | [330, 336] |
| Artificial Neural Networks (ANNs) | Tensor Flow | [434, 333, 376, 326] |
| Multi-Layer Perceptron (MLP) | Tensor Flow | [322, 319, 419, 326, 429, 342] |
| Convolutional Neural Network (CNN) | Tensor Flow | [290, 390, 374, 314] |
| Long Short-Term Memory (LSTM) | Tensor Flow | [342, 326] |

# 4 Discussion

Many studies have proposed that AI assessments for dementia are practical, noninvasive, user-friendly, and easy to be employed[327] and have potential to be integrated into dementia care settings. So, clinicians can leverage AI assessments to diagnose PwD or patients with subtypes of dementia [435, 373, 436]. However, due to complexity of AI assessments, encouraging the medical community to employ them is challenging. Thus, AI developers should consider developing explainable AI (XAI) assessments, which can describe their mission, rationale, and decision-making process underlying their results for clinicians. Thus, clinicians can gain insight into AI-made decisions [437]. Another significant barrier for physicians in adopting AI is the difficulties for convincing patients and their family members to accept AI-obtained clinical decisions. To overcome the barrier, clinicians and AI developers should cooperate in designing trustworthy AI assessments focusing on establishing ethical AI assessments that do not violate principles of autonomy, beneficence, non-maleficence, and justice [437, 438].

## 4.1 Challenges in Collecting and Sharing patients' data

Collecting patients' data is a multi-stage procedure that starts with a study design, recruiting participants following diversity and inclusion principles. Other stages are processing informed consent and evaluating the capacity of participants, in particular patients, to attend studies. Each step has its difficulties and concerns. For example, the sample size should be calculated for recruiting participants to show clinically significant results correctly [439]. If clinicians recruit fewer participants (patients and healthy controls), the data collection would be affordable and require much time. Furthermore, the delivery of services and the ability of PwD to operate the system provide a significant obstacle in collecting sensor data. We acknowledge that some PwD experience difficulties, notably in access and digital literacy, despite the potential of digital technologies. For instance, in research that examined the experiences of people who had received dementia care after being diagnosed, respondents thought that online forms of support were inappropriate and unavailable to those without internet or electronic device access [228]. Also, recent research exploratory studies indicate that individuals with cognitive impairment may encounter social difficulties online, such as stereotyping language and unfavorable remarks [440]. The pertinent design challenges, such as visual design, feedback, screen size, customization, and user-friendly controls, could provide another difficulty for them [441].

Sharing patients' data with other communities is essential because developing new AI assessments for dementia requires international multidisciplinary collaboration between multiple research centers and healthcare services. Furthermore, sharing patients' data is also necessary to improve research rigor, transparency, and replicability in dementia and AD [442]; Undoubtedly, patients' data are sensitive to be shared with unauthorized parties without considering privacy and confidentiality protocols. Thus, Sharing patients' data should be done following privacy principles.

## 4.2 Challenges in Developing, Deploying and Integrating AI Assessments for Dementia

Although there are many advantages of deploying AI assessments for dementia and integrating them into dementia care settings, there are certain challenges with deploying various AI assessments including voice-based assessments and sensor-based assessments. For example, sensor-based AI assessments may also present important ethical difficulties [443, 444], since they are being created for frail, elderly persons who frequently lack the mental capacity to provide their consent to their use. Moreover, information retrieval and data analysis may contain sensitive data including personally identifiable information like medical information or video recordings of an individual's behavior. As significant ethical considerations in the application of AI in PwD, informed permission, autonomy, privacy, data security, and affordability have also been noted in two prior assessments [445, 446]. Additional ethical issues covered in the literature include stigma, social exclusion, a lack of user involvement in technology development, access to assistive technologies, cost, use and the moral conundrum of whether such technologies should or should not take the place of human caregivers [228, 447, 448].

The main challenge for integrating AI into health care settings is the reluctance of the medical community to employ AI assessments to identify people with dementia. To foster clinicians' confidence in employing AI, AI developers should consider developing explainable AI (XAI) systems, which can describe their mission, rationale, and decision-making process underlying their results for clinicians; thus, clinicians can gain insight into the AI systems' decision-making processes [437].

Another significant barrier for physicians in integrating AI is convincing patients and their family members to accept results obtained by AI. To overcome the obstacle and encourage caregivers to get results obtained by AI systems, clinicians and AI developers should cooperate in designing trustworthy AI systems, focusing on establishing ethical AI systems that do not violate principles of autonomy, beneficence, non-maleficence, and justice [437]. Hence, it is thought to be a beneficial technique to predictably examine the practical, technological, clinical, and ethical difficulties connected with AI systems to look into the perspectives and demands of both patients and relevant stakeholders of technology development and healthcare [447]. A recent exploratory study investigated the perspectives of German dementia caregivers and found that there is a "knowledge gap" between developers and end users, which is likely to hinder adequate end user adoption [449]. In another study [447], the author interviewed healthcare workers in three European nations using a similar methodology, demonstrating the perceived importance of making sure user-centeredness is maintained when creating assistive technologies and AI systems. A second UK study examined the opinions and experiences of dementia patients, their caregivers, and general practitioners regarding the use of technologies in dementia care. Moreover, the expenses of assistive technologies are very high, and it is unclear when to utilize them to improve health outcomes [228]. In addition to protecting patient privacy, maintaining patient data confidentiality is essential for creating dependable relationships between clients and service providers, whereas the handling, management, and dissemination of patient data must be expressly shared by solution providers. Nevertheless, despite the barriers and obstacles in AI technologies for PwD, a recent study that evaluated technologies for behavioral monitoring [450] found that, despite noting significant privacy concerns, caregivers typically view the use of AI for the home surveillance of people with dementia to be ethically acceptable.

Various AI assessments can be developed by training ML or DL algorithms with multi-modal data such as speech and language data, sensor data, and GPS-based data [451]. Such AI assessments can identify older adults with associated behavioral, functional and language disorders associated with dementia [452]. Furthermore, AI developers can combine augmented reality (AR) and virtual reality (VR) techniques with behavioral, functional, language and speech data to design multi-modal AR/ VR AI assessments to identify patients with dementia. In addition to the approaches mentioned above, music-based assessment methods, which can evaluate musical recognition capability [453] or art based assessment methods [454] can be combined with AI systems not to identify individuals at risk of dementia but also to track the progress of the diseases.

# Author Contributions

- MP: Developed the structure of the manuscript. Prepared the initial draft of the manuscript. Authored the sections titled "Introduction," "Non-AI Assessments for Dementia," "Voice-based AI Assessments for Dementia," "Neuroimaging-based AI Assessments for Dementia," and "Discussion." Revised and refined the manuscript. Finalized the draft. Conducted revisions on the manuscript. Provided the final version of the draft.

- HG: Contributed to the section titled "Facial Expression to Develop AI Assessments for Dementia."

- VD: Contributed to two subsections titled "Wearable and Optical Sensor to Assess Gait Patterns" and "Digital Pen to Develop AI Assessments for Dementia."

- CM: Worked on developing and editing the section on "Neuroimaging-based AI Assessments for Dementia."

- IL, SN, and YK: Collaborated on the sections titled "Home and Environmental Sensors to Develop AI Assessments for Dementia," "Challenges in Collecting and Sharing Patients' Data," and "Challenges in Developing, Deploying, and Integrating AI Assessments for Dementia."